\documentstyle[floats,twocolumn,prl,aps]{revtex}

\input epsf
\input rotate

\begin{document}

\title{Self-organization in systems of self-propelled particles}

\author{Herbert Levine and
Wouter-Jan Rappel
}
\address{Dept. of Physics \\
University of California, San Diego \\ La Jolla, CA  92093-0319}
\author {Inon Cohen}
\address{ School of Physics and Astronomy\\
Raymond \& Beverly Sackler Faculty of Exact Sciences\\
Tel-Aviv University, Tel-Aviv 69978, Israel
}

\author{\parbox{397pt}{\vglue 0.3cm \small
We investigate a discrete model consisting 
of self-propelled particles that obey simple
interaction rules.
We show that this model can self-organize 
and exhibit coherent localized solutions 
in one- and in two-dimensions.
In one-dimension, the
self-organized solution is a localized flock of finite extent 
in which the density abruptly drops to zero at the edges. 
In two-dimensions, we focus on the vortex solution
in which the particles rotate around a common center 
and show that this solution can be obtained from random initial 
conditions, even in the absence of a confining boundary.
Furthermore, we develop a continuum version of our
discrete model and demonstrate 
that the agreement 
between the discrete and the continuum model is excellent.
}}

\maketitle

\vspace{2cm}

Self-organization and pattern formation in 
systems of self-propelled entities are ubiquitous in nature.
Examples can be found in a variety of fields and include
animal aggregation \cite{ani}, traffic patterns \cite{tra} and
cell migration \cite{cel}.
Recently, 
the problem of flocking, in which a large number of 
moving particles (e.g. fish or birds)  remain coherent over 
long times and distances, has attracted  considerable attention. 
A simulation of a simple numerical model by Vicsek {\it et al.}
\cite{vicetal}
in which each particle has a constant identical speed 
and a direction of motion that is determined by the average
direction of its neighbors,
revealed that an ordered phase exists, even in the present 
of noise and disorder.
A subsequent analysis of a continuum model  by Toner and Tu \cite{tontu}
investigated this ordered phase further
and derived conditions for its stability.

These models have in common that the flocks have 
infinite extent and, 
in simulations, fill the entire computational box. 
In reality, however, flocks have a finite size, with its density 
dropping sharply at the edge of the flock \cite{mogede}. 
In this Letter
we present a discrete model consisting of self-propelled interacting
particles that obey simple rules. 
We show that self-organization in 
our model leads to coherent {\it localized} states
in one dimension (1D) and in two dimensions (2D) that are stable 
in the presence of noise and disorder.
Furthermore, we present a continuum version of our discrete 
model which is obtained by coarse-grain averaging the discrete
equations. 
The continuum flock solutions in 1D agree very well with the discrete
solutions and are characterized by having
a finite extent with densities that
drop off sharply at the edges.
In 2D, we focus on a vortex state in which 
the particles are rotating around a common center and 
show that the discrete model and the continuum model agree 
well.

\begin{figure}[ht]
\def\epsfsize#1#2{0.3#1}
\newbox\boxtmp
\setbox\boxtmp=\hbox{\epsfbox{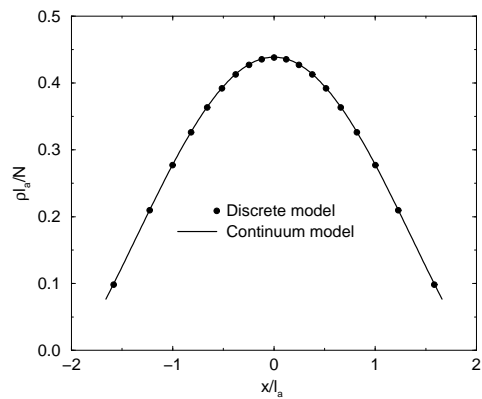}}
\rotr{\boxtmp}
\vspace{0.5cm}
\caption{
A coherent moving flock in the one-dimensional version of the model with
parameters (all with arbitrary units) 
$m$=1, $\alpha$=0.5, $\beta$=1,  $C_a$=0.45,
$l_a$=60, $C_r$=2 and $l_r$=20.
The solid circles correspond to the solution of the discrete model for
$N$=200 and every 10th particle displayed. 
The solid line shows the solution of the 
continuum model.
}
\label{1d}
\end{figure}

\begin{figure}[t]
\def\epsfsize#1#2{0.35#1}
\newbox\boxtmp
\setbox\boxtmp=\hbox{\epsfbox{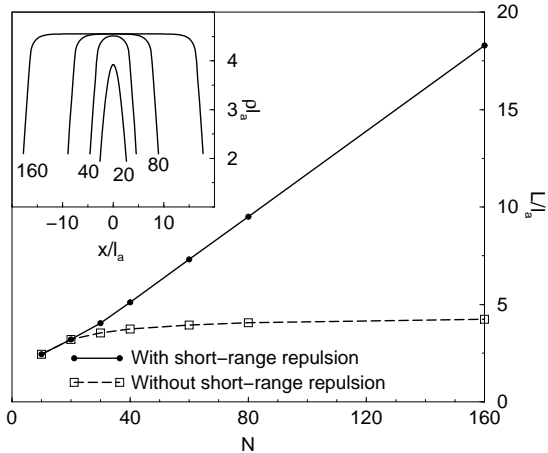}}
\rotr{\boxtmp}
\vspace{.5cm}
\caption{
The size $L$ of a flock as a function of $N$ with (solid line) and
without (dashed line) an additional short-range hard-core
potential.
The inset shows the density of flocks in the presence of
the short-range hard-core 
potential for different $N$. 
Parameter values: 
$m$=1, $\alpha$=0.5, $\beta$=1,  $C_a$=0.6,
$l_a$=40, $C_r$=2, $l_r$=20, $C_{hc}$=1 and
$l_{hc}$=10.
}
\label{hc}
\end{figure}
Our discrete particle model consists of $N$ particles with mass $m_i$,
position $\vec{x}_i$ and velocity $\vec{v}_i$.
Each particle experiences a self-propelling force $\vec{f_i}$
with fixed magnitude $\alpha$.
To prevent the particles from reaching large speeds, 
a friction force with coefficient $\beta$ is introduced.
In addition, each particle is subject to an attractive force which is
characterized by an interaction
range $l_a$.
This force is responsible for the aggregation of the particles and 
corresponds in animal aggregates 
to an awareness of the position of surrounding animals.
To prevent a collapse of the aggregate, a shorter-range repulsive force is
introduced with interaction
range $l_r$.
Thus, the governing equations for each particle is 
\begin{eqnarray}
m_i \partial_t \vec{v}_i
=
\alpha \hat{f_i} -\beta \vec{v}_i-\vec{\nabla} U 
\label{dis1}
\\
\partial_t \vec{x}_i=\vec{v}_i
\label{dis2}
\end{eqnarray}
We have checked that our qualitative results are 
independent of the explicit form of the interaction potential  and 
we have chosen here an exponentially decaying interaction:
\begin{eqnarray}
U= \sum_{j\ne i} C_a
\exp (- |\vec{x}_i-\vec{x}_j|/l_a) 
\nonumber \\
- \sum_{j\ne i} C_r
\exp (- |\vec{x}_i-\vec{x}_j|/l_r)
\end{eqnarray}
where $C_a$, $C_r$ determine the strength of the attractive and 
repulsive force respectively. The direction of the 
self-propelling force can be chosen along
the instantaneous velocity vector 
or, similar to the numerical model of Ref. \cite{vicetal},
can be determined by
aligning it with the average velocity direction
of the neighboring particles: 
\begin{eqnarray}
\vec{f}_i=\vec{v}_i & \ \ \ &{\rm without \ averaging}
\nonumber
\\
\vec{f}_i=\sum_{j\ne i} \vec{v}_j \exp (- |\vec{x}_i-\vec{x}_j|/l_c) 
& 
& {\rm with \ averaging} \\
\end{eqnarray}
where $l_c$ is a correlation length.

Let us now present our simulation results  
of the discrete model. 
The model was integrated by solving Eqs. \ref{dis1},\ref{dis2} 
using a simple Euler integration 
routine with timestep $\Delta t=0.2$. 
The simplest coherent {\it localized} 
solution in our model is a 1D
flock in which all particles move with constant 
velocity $v=\alpha/\beta$.
An example of this solution is presented in Fig. \ref{1d} 
where we have plotted, as solid circles, the density defined as
$\rho_{i}=2/(x(i+1)-x(i-1))$ as a function of the position of the 
particle. 
The density can be seen to drop abruptly to zero at the edge 
of the flock. 
We have checked that this solution is stable in the presence of 
moderate amounts of 
noise (added to Eq. \ref{dis2}) and of disorder in the 
parameters.

Further simulations revealed that increasing the number of 
particles does not change the shape of the density function
and that the total size of the flock reaches a constant 
value.
This is illustrated in Fig. \ref{hc} where the dashed line 
represents the size of the flock as a function of the 
number of particles. 
This obviously unrealistic behavior of the model is
due to the soft-core repulsive force which allows the 
particles to be very close. 
Our model can easily be extended to incorporate 
hard-core repulsive forces.
In fact,
a flock with a size that scales approximately linearly  with
the total number of particles can be obtained by 
introducing a hard-core repulsive force in Eq. \ref{dis1}
\cite{note2}.
The specific form of the hard-core  potential is not important 
and we have chosen 
\begin{eqnarray}
U_{hc}= 
\sum_{j\ne i} C_{hc} (|x_i-x_j|-l_{hc})^5 \ \ \ \ \ |x_i-x_j|\le l_{hc} 
\nonumber 
\\
U_{hc}= 0 \ \ \ \ \ |x_i-x_j|>l_{hc} 
\end{eqnarray}
In Fig. \ref{hc} we show, as a solid line, the
size of the flock in the presence of this additional repulsive force
as a function of $N$
while in the inset we show the corresponding density of the flock
for different $N$.
Naturally, the force has only an effect when the interparticle
separations are smaller than $l_{hc}$. Hence,
for small $N$  the flock solution is unaffected by the additional force.
As $N$ is increased and the interparticle spacing 
becomes smaller than $l_{hc}$, the particles in the center of 
the flock are pushed apart. 
For large $N$, the flock reaches a constant density in 
its center and its size scales linearly with 
$N$.

\begin{figure}[ht]
\epsfxsize=9.0cm
\epsfysize=8.0cm
\epsfbox{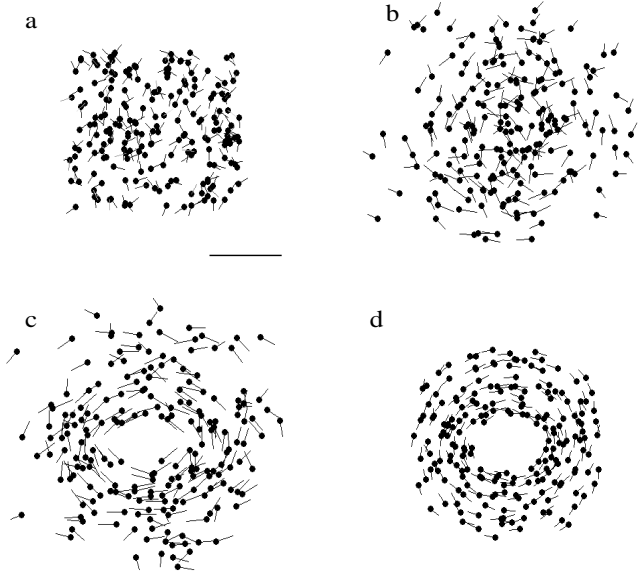}
\vspace{0.6cm}
\caption{
Snapshots of $N$=200 particles for the parameter values
$\alpha$=10, $\beta$=1, $l_c=0$,
$C_a$=0.4, $l_a$=40, $C_r$=1 and $l_r$=20.
As initial condition, the particles are placed at random on 
a disk with velocities that are constant in 
magnitude ($\alpha/\beta$) but random in direction (a). 
After an initial transient (b, 20 iterations and c, 
50 iterations), a stable rotating vortex 
state is formed (d, 300 iterations). 
The bar indicates the attraction length $l_a$.
The position of each particle is denoted by a solid circle and the 
velocity as a line starting at the particle and pointing in the 
direction of the velocity. 
}
\label{ic}
\end{figure}

Let us now turn to 2D where we have obtained several
flocking states. One, not shown here,  is the equivalent of 
our 1D flock: all particles are moving in the same direction 
with $|\vec{v_i}|=\alpha/\beta$.
The particles arrange themselves on a disk and this aggregate 
is stable under small amounts of noise and disorder.
The solution which we will focus on here
consists of a vortex state where the particles
rotate around a common center and which is common in
fish schools \cite{schools}, bacterial 
colonies \cite{bj} and amoeba aggregates \cite{cellagg}.
This solution has been observed previously  in 
models of self-propelling particles but only 
in the presence of a confining boundary \cite{dupetal,hem}
or when a rotational chemotaxis term is invoked \cite{bj}.

\begin{figure}[ht]
\def\epsfsize#1#2{0.35#1}
\newbox\boxtmp
\setbox\boxtmp=\hbox{\epsfbox{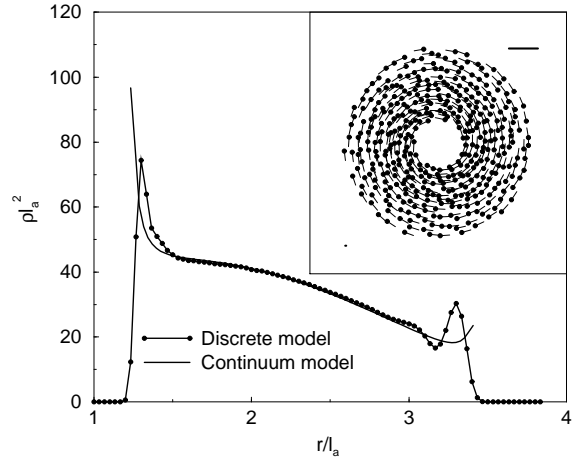}}
\rotr{\boxtmp}
\vspace{.5cm}
\caption{
Average density of a rotating vortex state 
in the discrete model (solid symbols)
and the continuum model for the parameter values $N$=400,
$m$=1, $\alpha$=10, $\beta$=1, $l_c=0$,
$C_a$=0.5, $l_a$=30, $C_r$=1 and $l_r$=20.
The inset shows
a snapshot of the discrete model simulation with the bar corresponding 
to $l_a$.
As initial condition we used a vortex obtained with $l_c=4$ which 
ensured that the angular velocity of all particles has the same sign.
}
\label{vortex}
\end{figure}

In our model,
the vortex solution can be obtained from a wide variety of initial
conditions
including one in which all particles are randomly placed on a
disk with speed $\alpha/\beta$ and random initial velocities.
A typical evolution of
the particles starting from this initial condition
is shown in Fig. \ref{ic}.
This figure was obtained in the absence
of the velocity averaging term and
illustrates that in this case some particles move clockwise while the
others rotate counter clockwise.
When the velocity averaging term is included the final vortex state
consists of all particles turning  the same way.
An example of this case is shown in the inset of
Fig. \ref{vortex}.
In both cases,
the speed of each particle was found to be sharply peaked around
$|\vec{v}|=\alpha/\beta$

The average size of the vortex remains constant in time
as shown in Fig. \ref{vortex} where we have plotted the average
density (obtained by averaging over $10^6$ iterations) of the
vortex structure.
Fig. \ref{vortex} displays several remarkable features. First,
there is a well-defined core which remains void even for extended
simulation runs. Using different parameter values however,
one can also produce a vortex without a core.
Second, as in the one dimensional
flock, the density does not decay smoothly to zero 
at the edges.
Instead, it increases at both the inner and outer edge
of the aggregate and then drops abruptly.
Qualitatively similar vortex solutions were found when 
an additional hard-core repulsion like the one discussed above
is added. 

As in the case of traffic models \cite{helbing}, it is useful 
to develop a continuum version of our model.
To this end, we simply coarse-grain average the equations which 
results in,  after dividing by a common factor of
$\rho$,
\begin{eqnarray}
\partial_t \vec{v}+(\vec{v} \cdot  \vec{\nabla}) 
\vec{v}
=
\alpha \hat{f}-\beta \vec{v} + \vec{G}
\label{cont}
\end{eqnarray}
together with the conservation equation for the density $\rho$
\begin{eqnarray}
\partial_t \rho+ \vec{\nabla} \cdot (\vec{v} \rho)=0.
\nonumber
\end{eqnarray}
The interaction force is given by
\begin{eqnarray}
\vec{G}(\vec{x}) 
= \int \rho(\vec{x}') \vec{\nabla} U(\vec{x},\vec{x}') d\vec{x}'
\end{eqnarray}
and the self-propulsive force direction is 
given by either
\begin{eqnarray}
\vec{f}(\vec{x}) = \int \rho(\vec{x}') \vec{v}(\vec{x}') exp(
-|\vec{x}-\vec{x}'|/l_d)d\vec{x}'
\end{eqnarray}
in the velocity averaging case or simply by
$\hat{f}=\vec{v}/|\vec{v}|$ otherwise.

\begin{figure}[ht]
\null
\vspace{-2.5cm}
\def\epsfsize#1#2{0.38#1}
\newbox\boxtmp
\setbox\boxtmp=\hbox{\epsfbox{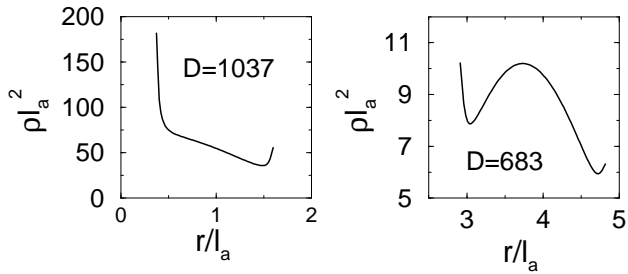}}
\rotr{\boxtmp}
\vspace{.5cm}
\caption{
Two different solutions of the continuum model for the parameters
$\alpha/\beta$=10,
$C_a$=0.7, $l_a$=40, $C_r$=2, $l_r$=20 and $N$=200.
}
\label{solutions}
\end{figure}

A comparison between the discrete model and the 
continuum model can be carried out for the solutions presented here. 
For a 1D flock, we have simply $\hat{f_i}=1$, and
the solution in the continuum model
is given by $G=0$ or
\begin{equation}
\int \rho(x') U(x,x') dx'=D
\label{1dcont}
\end{equation}
where
$D$ is a constant determining the total number of particles.
Since the sought after solution
has a finite extent with a discontinuity at the edge where the 
density drops to zero we discretize the integral using 
$M$ points and discretization step $\Delta x$.
The last point corresponds to the edge of the flock. 
The resulting linear set of $M$
equations for $\rho$ is easily solved using standard linear algebra
packages.  and $\Delta x$ was varied  until
the slope
at the center of the flock vanished.
The result, with $D$ chosen such that
$\int \rho(x) dx=N$, is shown in Fig. \ref{1d} as a solid line.
The density profile in the continuum model 
is discontinuous at the edge and agrees well with 
the profile obtained in the discrete model.
Note that since the equations are linear in $\rho$
it is not surprising to find that the density profile 
of the discrete model did not change as the number of particles
is increased.
Clearly,the simple coarse-grained averaging 
procedure is not adequate for the hard-core potential case,
where higher order terms in the density are important.

The continuum model can also be used to find the vortex solution.
To this end, we use the fact that all particles undergo approximately
a circular motion with constant speed ($\alpha/\beta$). 
Thus,
a continuum vortex solution can be found by requiring 
that the force $\vec{G}$ is centripetal:
\begin{equation}
\int_0^{2 \pi} d\phi \int_0^{\infty}
dr \rho(r) U(r,\phi) =D
-(\alpha/\beta)^2 {\rm ln} (r) 
\label{2dcont}
\end{equation}
After performing the integration over $\phi$, the 
remaining integral was discretized as in the 1D case.
The resulting
matrix was solved for  $\rho(r)$ and used in a Newton solver
that searched for the size of the hole and the overall size of the
vortex (i.e. discretization $\Delta x$)
with a condition for smoothness of the solution at
both discontuous edges. 
This condition simply consisted  of requiring that the first and last 
point can be obtained by linear interpolation from its two 
neighboring points. 
We have checked that the solutions we obtained are converged by 
increasing the number of discretization points from 80 to 1480.
In Fig. \ref{vortex} we compare the discrete solution to the 
one obtained by Eq. \ref{2dcont} where
the integration constant
$D$  was varied until $\int \rho(\vec{x}) d\vec{x}=N$.
Again, the continuum profile is discontinuous 
at the edges and  the agreement between the continuum profile 
and the discrete profile is very good.

The continuum equation can be used to explore the  (large)
parameter space more efficiently.
An example of such an exploration
is shown in Fig. \ref{solutions} where we plot two different
solutions found by our Newton solver for the same model parameters and 
total number of particles ($N$=200) 
but with different integration constant $D$. 
Preliminary simulations of the discrete model
show that the solution with the 
larger core is unstable and a formal stability analysis 
of the continuum solution will be carried out in the future. 

In this Letter we have presented a model for 
localized aggregates and flocks. Our model 
contains very simple rules and can be straightforwardly 
extended to 
incorporate additional  and different types of interactions. 
For example, the forces that maintain bacterial aggregates 
are believed to be short-range adhesion forces together with 
a short-range hard-core repulsion.
The investigation of these types of interactions will be the 
subject of future work. 
Finally, it would be interesting to compare our results 
to animal flocks. Unfortunately, such a comparison is 
currently not possible since not enough quantitative data
is available.

We acknowledge useful conversations with E. Ben-Jacob and 
W.F. Loomis.
The work of HL and WJR was supported in
part by NSF DBI-95-12809. IC acknowledges support from 
The Colton Scholarships and a 
Israeli-US Binational Science Foundation BSF grant.


\begin{thebibliography}{20}
\bibitem{note2} It is of course also possible to eliminate
the soft-core repulsive potential altogether. 
The soft-core repulsive potential is introduced here to 
facilitate a simple derivation of a continuum model.
\end{thebibliography}
\end{document}